\documentclass[10pt,conference]{IEEEtran}
\IEEEoverridecommandlockouts

\makeatletter
\renewcommand{\@IEEEsectpunct}{ \ \,}%
\makeatother

\usepackage{amsmath,amssymb,amsfonts}
\usepackage{algorithmic}
\usepackage{graphicx}
\usepackage{textcomp}
\usepackage{xcolor}
\def\BibTeX{{\rm B\kern-.05em{\sc i\kern-.025em b}\kern-.08em
		T\kern-.1667em\lower.7ex\hbox{E}\kern-.125emX}}

\usepackage{silence}

\usepackage{booktabs} %
\usepackage{microtype}
\usepackage[hyphens]{url}
\usepackage[numbers]{natbib}
\usepackage[moderate]{savetrees}
\usepackage{tabulary}
\usepackage[utf8]{inputenc}
\usepackage{placeins}
\usepackage[english]{babel}
\usepackage{csquotes}
\usepackage{xcolor}

\begin{document}
	
\IEEEoverridecommandlockouts
\IEEEpubid{\begin{minipage}{\textwidth}\ \\[10pt]
		\centering\footnotesize{\linebreak \linebreak \linebreak \linebreak \copyright ACM 2019. This is the author's version of the work. It is posted here for your personal use. Not for redistribution. The definitive Version of Record will be published in Proceedings of the 41st International Conference on Software Engineering. Please cite the paper as follows: Knutas, A., Palacin, V., Maccani, G., Helfert, M. 2019. Software Engineering in Civic Tech: A Case Study about Code for Ireland. In Proceedings of the 41st International Conference on Software Engineering. }
\end{minipage}}

\title{Software Engineering in Civic Tech \linebreak A Case Study about Code for Ireland}

\author{\IEEEauthorblockN{Antti Knutas}
	\IEEEauthorblockA{\textit{LUT University} \\
		\textit{School of Engineering Sciences}\\
		Lappeenranta, Finland \\
		antti.knutas@lut.fi}
	\and
	\IEEEauthorblockN{Victoria Palacin}
	\IEEEauthorblockA{\textit{LUT University} \\
		\textit{School of Engineering Sciences}\\
		Lappeenranta, Finland \\
		victoria.palacin@lut.fi}
	\and
	\IEEEauthorblockN{Giovanni Maccani}
	\IEEEauthorblockA{\textit{Maynooth University} \\
		\textit{School of Business}\\
		Maynooth, Ireland \\
		giovanni.maccani@mu.ie}
	\and
	\IEEEauthorblockN{Markus Helfert}
	\IEEEauthorblockA{\textit{Dublin City University} \\
		\textit{School of Computing}\\
		Dublin, Ireland \\
		markus.helfert@dcu.ie}
}

\maketitle

\begin{abstract}
Civic grassroots have proven their ability to create useful and scalable software that addresses pressing social needs. Although software engineering plays a fundamental role in the process of creating civic technology, academic literature that analyses the software development processes of civic tech grassroots is scarce. This paper aims to advance the understanding of how civic grassroots tackle the different activities in their software development processes. In this study, we followed the formation of two projects in a civic tech group (Code for Ireland) seeking to understand how their development processes evolved over time, and how the group carried out their work in creating new technology. Our preliminary findings show that such groups are capable of setting up systematic software engineering processes that address software specification, development, validation, and evolution. While they were able to deliver software according to self-specified quality standards, the group has challenges in requirements specification, stakeholder engagement, and reorienting from development to product delivery. Software engineering methods and tools can effectively support the future of civic technologies and potentially improve their management, quality, and durability.
\end{abstract}


\section{Introduction}

\begin{figure}[!ht]
    \centering
    \includegraphics[width=2in]{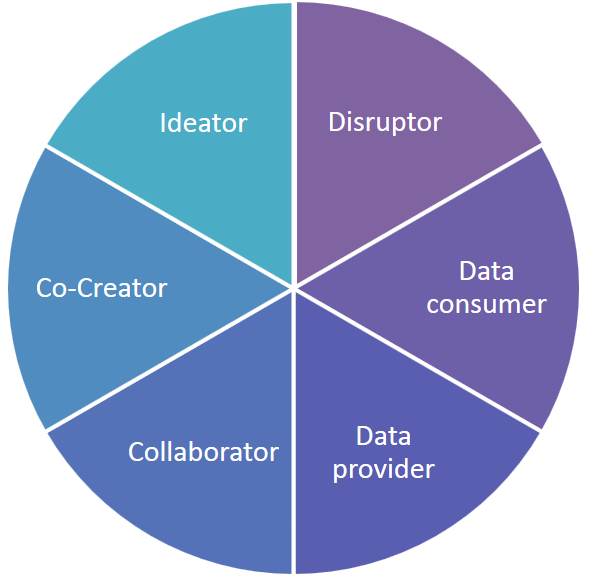}
    \caption{Palette of Public Participation in Civic Technology adapted from \cite{palacin2018ict4s}}
    \label{fig:palette}
\end{figure}

Technology has facilitated the evolution of public participation. It has enhanced the rise of civic grass rooted actions such as activism, mobilizations, public campaigning and community monitoring due to its power to connect people, improve cities and to better governance \cite{KnightFoundation2013}. It must be understood that every interaction between people and a civic technology\footnote{Civic technology refers to the diverse ways in which people are using technology to influence change in society \cite{KnightFoundation2013,Steinberg2014,boehner2016}.} represents a deliberate and intentional act of public participation, which can take different forms (see Fig.~\ref{fig:palette}), from merely allowing people to consume information about matters they care about e.g. traffic or pollution (data consumer); collect data about predefined issues of common interest e.g. FixMyStreet (data provider); collaborate with authorities to monitor issues - predefined by authorities - (collaborators); co-create solutions for issues of shared concern (co-creator); ideate civic action e.g. Code for Ireland (ideator); or, disrupt established processes by passive non-participation or negative participation (disruptor) \cite{palacin2018ict4s}.

Civic technology is currently an intense target of study \cite{saldivar2018civic}. While extant literature seeks to understand the effect of technology on people \cite{boehner2016,johnson2016} and to orchestrate civic engagement \cite{ferrario2014software, balestrini2017city}, research in the context of software engineering practices in civic technologies remains under-explored \cite{gama2017preliminary,lee2015open}.

In order to better understand how grassroots organizations develop civic tech, we performed an in-depth qualitative study of a civic tech group: Code for Ireland. In the study, we followed the formation of two projects in Code for Ireland, seeking to understand how their development processes evolved over time, and how the group carried out their work in creating new technology. We want to understand if a systematic process similar to a software engineering process emerges, and what kind of process the group adopts. From existing studies on open source software projects \cite{scacchi2006understanding}, we know that volunteer-based communities can set up processes with characteristics of software engineering processes, including the systematic use of theories, methods, and tools to achieve desired levels of quality, acceptability, and maintainability \cite{sommerville_software_2011}.

To accomplish our goals, we framed our research as an interpretive case study \cite{walsham1995interpretive, runeson2009guidelines} using the constructive grounded theory research method \cite{Charmaz2014Constructing}\footnote{In this context theory is defined as a set of testable abstract statements that define and explain relationships among constructs, as set of prescriptive instructions on how something should be done, and as a set of abstract statements providing a lens for viewing or explaining a part of a phenomenon \cite{gregor2006nature}. A situational theory emerging from the research process is not necessarily a "grand theory" explaining the entire phenomenon \cite{Urquhart2010Putting}.}. Our main contribution is providing a situated, systematic explanation of one civic technology software engineering process phenomenon.

To address this gap in knowledge, we have formulated the following exploratory research question for our study: \textit{What software creation processes emerge in grassroots-driven civic technology groups?} As secondary results 1) we also relate our results to established literature on software engineering and civic technology processes, and 2) provide actionable knowledge in the form of recommendations to emerging grassroots civic tech organizations using lessons learned and established literature.

The paper is structured as follows. In the next section, we perform a literature review on related civic technology studies. In section three we detail the research setup and in section four our analysis and findings. We discuss the findings in section four and conclude with section five.

\section{Related Research on Civic Technology}

Since 2014, the term civic technology, also known as "civic tech", started to increasingly appear in discussions about public participation and governance, rapidly surpassing terms like e-governance, e-democracy, gov2.0 or gov.3.0 \cite{Steinberg2014}. Civic tech is a term that refers to the diverse ways in which people are using technology to influence change in society \cite{KnightFoundation2013}. The breadth of civic technologies is wide and comprises a large pool of technologies for i) governance (e.g. MySociety, SeeClickFix), ii) collaborative consumption (e.g. Airbnb, TaskRabbit), iii) community action (e.g. citizen investor, GeekCorps), iv) civic media (e.g. Wikipedia, Global Voices) and v) community organizing (e.g. Whatsapp groups) \cite{KnightFoundation2013}.

Researchers across many fields of software engineering have been studying civic technologies from different perspectives, for instance: a) examining the role of information technologies in governance \cite{borg2018digitalization}; b) supporting participatory democracy practices of technology in society \cite{holston2016engineering}; c) understanding the role of software in society \cite{newman2015role}; d) exploring the dynamics of the collaborative use of software \cite{franzago2017collaborative}, e) developing agile methods and techniques to support software development in society \cite{memmel2007agile,gama2017preliminary} and, more recently f) architecting smart cities, open data with and for people \cite{larrucea2017software}. In other related fields, such as HCI, the focus has turned onto examining the a) dynamics of technology in public life, b) civic discourse on social media, c) civic engagement, d) data literacy, e) social justice and f) trust, often using in-the-wild approaches of systems in communities \cite{boehner2016,johnson2016}. In addition, both SE \cite{becker2015sustainability,penzenstadler2018software,penzenstadler2017software} and human-computer interaction (HCI) \cite{disalvo2010mapping,silberman2014next,knowles2018changes} share a recent turn of interest onto sustainability matters, which has a close link with civic tech, as the focus is put on the environment, peoples' behavior and the role of technology \cite{boehner2016}.

People have influenced the development of cities around the world by orchestrating grassroots initiatives through which they take action and solve issues they care about. These groups have affected positive change in neighborhoods as well as at a larger scale. For instance, Safecast\footnote{\url{https://blog.safecast.org}} was created by a group of people that wanted accurate and real-time measures of radiation. They created open hardware devices to allow people to measure radiation on the go. By now, Safecast has become the go-to solution for radiation measures worldwide (all supported by a large pool of volunteers across the globe). Civic technology groups have a rich variety of goals, from city-driven projects \cite{hou2017sustainable}, issue-centric initiatives \cite{balestrini2017city} to user-driven groups, such as hackerspaces and fablabs \cite{capdevila2014can}. Different ways to arrange public participation in the creation of civic tech have been developed, such as the City in Common framework \cite{balestrini2017city}, the MK:Smart project \cite{gooch2018amplifying}, and Code for Europe-style open innovation intermediates \cite{almirall2014open}.

\section{Research Setup}

The purpose of this research project is to investigate the software creation processes in grassroots-driven groups who create civic technology, about which our literature review informs a substantial lack of theoretical insights specifically related to this recent academic conversation. We assume that grassroots civic software creation contexts happen through an apparently natural process, but a complex one, and contingent on several social actors and activities. Therefore, we consider an interpretive perspective the most suitable for this project.

Given the nature of this study, we found a qualitative approach to data collection and analysis appropriate. Given that these software creation processes are explored from the meaning given by the people that are actively involved in constructing such reality, qualitative research methods are appropriate as they are “designed to help researchers understand people and the social and cultural contexts within which they live” \cite{Myers1997Qualitative}.

Case study research was found to be the most suitable for the purposes of this research. It is a qualitative approach in which the investigator explores a bounded system (a case in a specific setting/context) over time, through detailed, in-depth data collection involving multiple sources of information, and reports a case description and case-based themes \cite{walsham1995interpretive,yin2009case,Runeson2012Case}.

\subsection{Case: Code for Ireland}

The case we studied is concentrated on a grassroots group creating civic tech, Code for Ireland, which describes its mission as \textit{"developing innovative and sustainable solutions to real-world problems faced by communities across Ireland, by fostering collaboration with civic-minded individuals, businesses and public sector organizations."} The group is led by volunteers, coordinated online through services such as Meetup.com, and not officially registered. We coined a term \textit{"organically grown civic tech"} to describe the group because the current iteration of Code for Ireland was formed by a group of motivated citizens coming together. They have not registered as a formal organization, they operate without the intention of making a profit, and use only volunteered or donated resources. Furthermore, there is no city organization, academic body, or other influential stakeholders backing the group. If evaluated with the palette of public participation (Fig.~\ref{fig:palette}), they are ideators or people who are taking civic actions in their hands and orchestrating the creation of civic tech.

Code for Ireland was selected as the case study for this research for several reasons, including: (1) the case is a clear example of grassroots civic software development (i.e. the phenomenon of interest); (2) since the very first contact with the case, a significant interest was observed from the group members to take part of this study, which ensured us the possibility of conducting interviews; (3) their willingness to share experiences, and access to relevant documents and material; (4) the group and its people are based in a location that allowed us long-term engagement and to conduct observations; and (5) unlike similar groups, Code for Ireland has been operating for more than 5 years, and thus ensured maturity in relation to the richness of the participants’ experiences. 

\begin{figure} 
    \centering
    \includegraphics[width=\columnwidth]{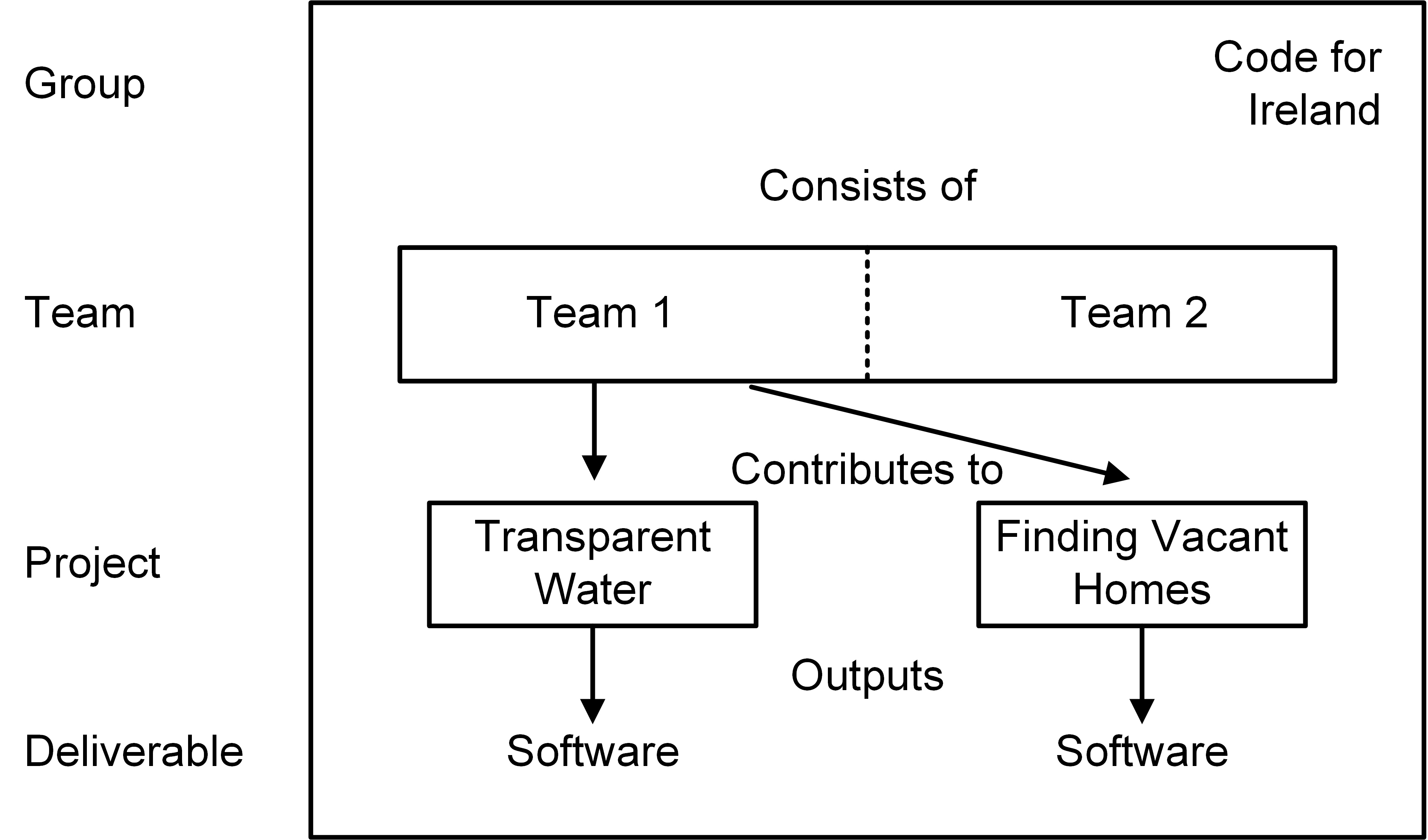}
    \caption{Case study context}
    \label{fig:researchcontext}
\end{figure}

The structure of the case study is described in Fig.~\ref{fig:researchcontext}, which is similar to a single case study with multiple embedded units of analysis \cite{Runeson2012Case}. In the figure, we specify three essential units: (1) The \textit{group}, which is all of Code for Ireland. (2) \textit{Teams}, who work on (3) \textit{projects}. The group can have several ongoing projects, with each team working on one or several projects simultaneously. In this case, we studied a single organization and the two projects the group was working on. These are Transparent Water, which is concentrated on making water quality data available; and Vacant Homes, which focuses on developing a mobile app to identify apartments that could be rented to alleviate the current housing crisis in Dublin. The projects are expanded on in Table~\ref{tab:c4iprojects}.

\begin{table}
\caption{Code for Ireland projects included in the study}
\label{tab:c4iprojects}
\def\arraystretch{1.5}
\begin{tabulary}{0.9\columnwidth}{p{1.1cm}LLL}
\toprule
\textbf{Project} & \textbf{Participants} &  \textbf{Stage} & \textbf{Description}
\\ \hline
Transparent Water & 4 core team; 1 mentor; 1-2 others & Achieved minimum viable product; preparing for a beta release. & The project aims to open up water quality data from Irish Water with an app. Main features include geolocation-based data summaries and service notifications.
\\ \hline
Identifying Vacant Homes & Fluctuating; 2-3 (core team); 1 mentor & Undergoing a restart after a switch of development team. & Collecting information on vacant homes through a mobile app and then providing summaries of this data to the official vacant home project.
\\ \bottomrule
\end{tabulary}
\end{table}

\subsection{Data Collection}

Inductive qualitative case study researchers usually combine multiple data collection methods \cite{walsham1995interpretive,eisenhardt1989building,eisenhardt2007theory,stake2013multiple} and keep the design of the process flexible to “provide stronger substantiation of constructs” \cite[pp. 538]{eisenhardt1989building}. While observation was a natural source for collecting data during the time we spent on site, documents and semi-structured interviews were chosen as the other main sources for the data collection process.

The research and data collection occurred during 2018 over a nine-month period. We interviewed\footnote{Interview guide available at https://doi.org/10.5281/zenodo.2565435} six participants from two projects. Additionally, a researcher participated in the activities of the group, including weekly online stand-ups and monthly in-person meet-ups for a total of 40 attended events, with events lasting from one to three hours on average. This involvement allowed exclusive access to the digital cooperation platforms of the group. Data from interviews was transcribed on the same day in which these were undertaken. Data from observation was stored in field notes. 

In case study research, consideration must be given to construct validity, internal validity, external validity, and reliability \cite{yin2009case,stake2013multiple}. We ensured construct validity through a validation effort conducted by presenting the findings back to the participants, and through building a chain of evidence throughout the coding process. Internal validity was ensured by including more than one researcher in the coding process.

\section{Case Analysis and Findings}

In this section, we present our data analysis process and summarize results from each stage of analysis. We analyzed the data using the grounded theory research method \cite{Glaser1978Theoretical,Urquhart2010Putting,Charmaz2014Constructing}, following the existing empirical software engineering literature in adapting it for case study analysis \cite{Runeson2012Case} and used e.g. by Kasurinen et al. \cite{kasurinen2010test}. Grounded theory has been useful in developing context-based, process-oriented descriptions and explanations of phenomena \cite{Urquhart2010Putting}.

There are variations on how to approach grounded theory research, as evidenced by a split between Strauss and Glaser in their research methodology publications. In this paper, we follow Charmaz's \cite{Charmaz2014Constructing} constructivist grounded theory, which acknowledges that when interacting and interviewing, knowledge is built together with the participants and the researcher. This approach was chosen because the researcher spent an extended period interacting with the researched civic tech group. We also used Charmaz's guidelines for ethnographic research for analyzing the data from research participation in Code for Ireland activities and the approach by Vaast et al. \cite{Vaast2017Building} when analyzing publicly available social media and online software repository data.

The grounded theory approach \cite{Charmaz2014Constructing} we selected has three stages, with each stage leading towards the increasing depth of analysis and generalization. The three stages \cite{Urquhart2010Putting} are summarized in Table~\ref{tab:gtlevels} and their results are expanded on in the following subsections.

\begin{table}
\caption{Three major steps of grounded theory research with increasing depth of analysis \protect\cite{Urquhart2010Putting}}
\label{tab:gtlevels}
\def\arraystretch{1.5}
\begin{tabulary}{\columnwidth}{LL}
\toprule
Open coding (description) & Describing conceptual constructs involved in the phenomenon and their properties.
\\ \hline
Selective coding (interpretation) & Defining and explaining the interactions between the conceptual constructs. Refining and generalizing. Understanding and explaining the area under investigation.
\\ \hline
Theoretical coding (formulating a theory) & Formulation of a descriptive theory. Aim is to create inferential and/or predictive statements about the phenomena. Achieved by defining relationships between individual interpretive constructs. E.g. associations, influences or causal.
\\ \bottomrule
\end{tabulary}
\end{table}

\subsection{Open Coding: Identifying Actors, Concepts and Categories}

We started the open coding phase by coding the interviews line by line using the Dedoose Computer Assisted Qualitative Data Analysis Software. The start of our coding process was informed by the \textit{process} coding family \cite{Glaser1978Theoretical}, which was selected as it best matched our research questions. With this open coding process we identified initial concepts present in the interviews and online material and defined 372 codes to describe them. During subsequent iterations these codes were grouped into eleven top categories, which are \textit{pre-process}\footnote{When we discuss a concept identified through the coding process, the word is italicized.}, \textit{team formation \& selling the idea}, \textit{group goals / decisions}, \textit{personal motivations}, \textit{challenges}, \textit{problem / opportunity}, \textit{requirements definition}, \textit{task division \& decisionmaking}, \textit{design \& engineering activities}, \textit{evaluation}, and \textit{post-process}.

When analyzing the interview data, we identified four main type of participants: \textit{Mentor}, \textit{core group}, a \textit{regular contributor}, and \textit{visitors}. Mentor spars with the group without directly influencing the decision-making process. The core group emerges from regular contributors by their activity and commitment, and has the most influence, mainly because they are the major contributors. If a \textit{project manager} is selected, the project manager is first among equals, others willingly following the project manager's lead. Most interviewed core group members and contributors had a software technology career background, and all were from technical fields, such as statistics and project management. None of the participants had previously participated in the open source software community, though some had worked with open data initiatives.
\begin{displayquote}
``[...] but also when people were committed, they were motivated that they were part of something like task or group that like solving real challenging problems.'' ---Interviewee D4
\end{displayquote}

During the last iteration of open coding, we selected the \textit{civic tech software creation process} as our core category, which is the central phenomenon around which all the other categories are related \cite{Strauss1990Basics}. In the next part, selective coding, we relate other discovered code categories to the core category.

\subsection{Selective Coding: Identifying the Civic Tech Software Development Process}

\begin{figure}
    \centering
    \includegraphics[width=\columnwidth]{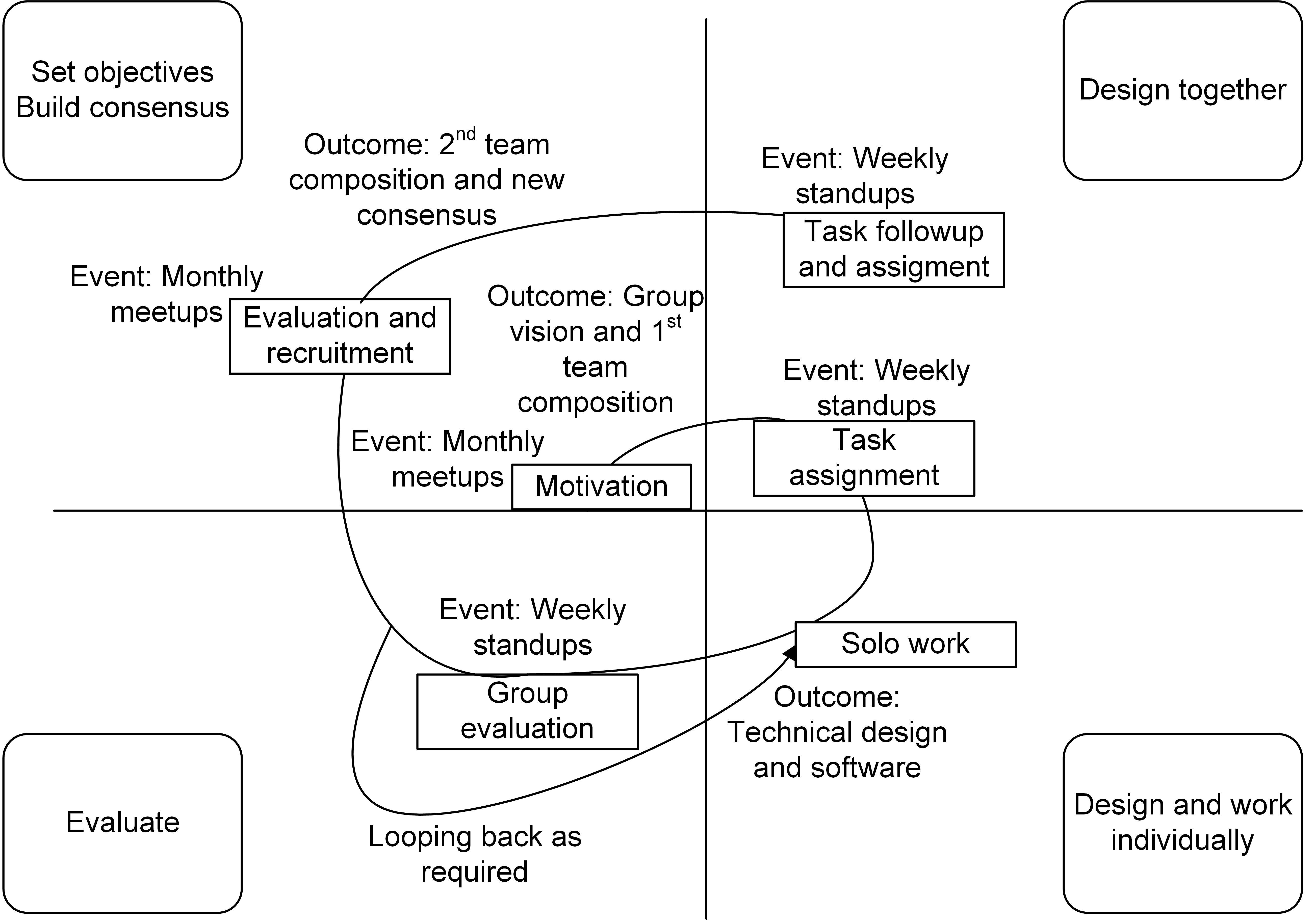}
    \caption{Overall project progress structure in Code for Ireland}
    \label{fig:workloop}
\end{figure}

In this section, we refine the data by taking the discovered codes, further categorizing them into hypernyms with the help of the \textit{process} coding family and then establishing causal or semantic relationships \cite{spradley_ethnographic_1979} between the concepts. We started selective coding by identifying the iterative process of the group's project structure and abstracting it. It is presented in Fig.~\ref{fig:workloop} and consists of \textit{monthly meetups}, which lead to \textit{new team composition}, who meet \textit{weekly to design} or to \textit{evaluate outputs}. They are guided by a \textit{mentor} in this process, until the \textit{new team} selects a \textit{project manager} or a \textit{core group} emerges. In one sense the process is iterative, but we have chosen to depict it as a spiral because some stages are not repeated. Furthermore, the project can change shape and direction rapidly if necessary and so desired by the team.

\paragraph*{Monthly meetups} are the main structured social event around which new teams and projects form. The meet-ups allow evaluating new ideas and getting input in the form of new members, ideas, and priorities. \textit{Motivation} and the \textit{selling the idea} are big part of the initial meet-up. The \textit{mentor} leads the first meet-up and presents projects to newcomers. The organization, Code for Ireland in this case, often has some ongoing projects or new ones waiting for implementation. The \textit{mentor} tries to make sure that individual goals match with the available projects and then facilitates the newly formed team's decision-making process until they create their own internal structure.
\begin{displayquote}
    ``I would say they definitely start mostly with the meet-up. They would come to a meet-up, get to know us during the meet-up and say, `Okay, I’m happy I’m interested in working a lot, continue working on it.' In some free time I’ll have just-- I don’t know if I have a lot of free time but I’ll see how much I can do and then go from there.'' ---Interviewee D5
\end{displayquote}

The first \textit{monthly meetup} results in initial consensus on \textit{group goals}, which is based on negotiation between team members and their \textit{motivations}. \textit{Weekly meetups} are the second major event around which the teams are structured. The meetups and project structure are loosely inspired by \textit{agile methodology}. The team members negotiate between task assignment before dispersing to perform individual work.
\begin{displayquote}
    ``It's an agile approach and -- if the team consensus says to go down a different direction, that's the direction it goes down.'' ---Interviewee D3
\end{displayquote}

\begin{figure}
    \centering
    \includegraphics[width=\columnwidth]{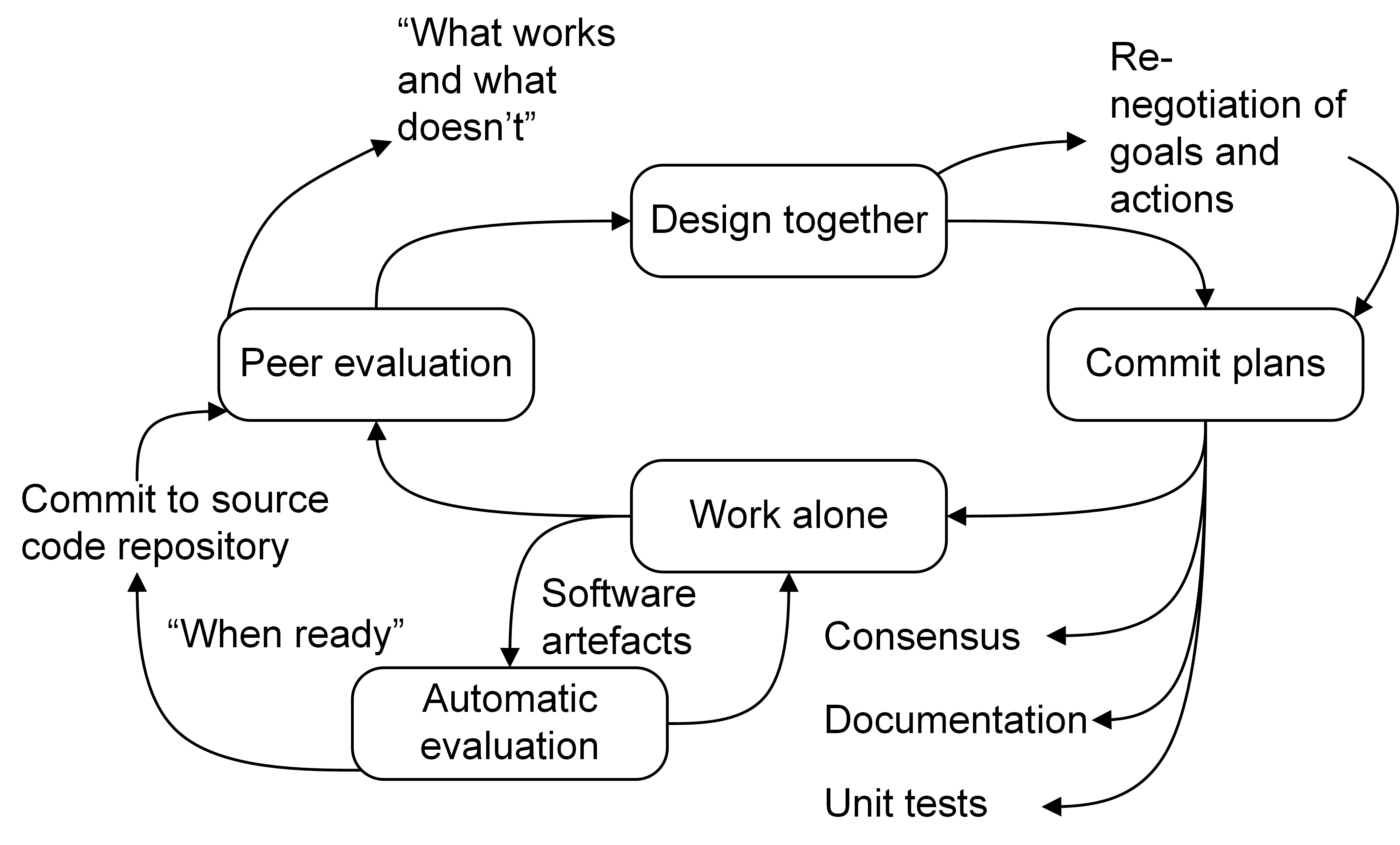}
    \caption{The teams' iterative plan-consensus-action-evaluation loop}
    \label{fig:designloop}
\end{figure}

\paragraph*{Weekly online stand-up meetings} and \textit{individual work} balance the team decision-making process and getting things done, and establish the \text{basic loop} of work. The teams meet weekly or less often (but more often than the monthly personal meeting) to \textit{design} and \textit{evaluate}. After the weekly meeting, members disperse into smaller units, like pairs or individuals, and work to accomplish the agreed on \textit{goals}. The loop between shared design and individual work and the outputs of each stage are depicted in Fig.~\ref{fig:designloop}.
\begin{displayquote}
    ``--We do like a week-by-week. We have a weekly review call where we talk about, 'Okay, this is where we're currently at, this is how much, what everyone's done.' Everyone does a stand up in a way and so what they've accomplished in the last week or 10 days and then we go from there, 'Okay, see, this is what we need to go forward, this is our current status it might depend on hosting, what technologies do we still need, what could be added, what functionality, where do we go from there?' '' ---Interviewee D5
\end{displayquote}

One major characteristic of this loop is \textit{opportunistic design} and \textit{opportunistic evaluation}. The goals are not only evaluated based on what needs to be done based on requirements but what can be done. This is because the group is dependent on volunteers and voluntarily contributed resources, such as server space or meeting locations. For example, if a team gains a member with competence in a new beneficial technology, the use of that technology might be re-negotiated into team goals. The team also recognizes that some features might not meet original goals, but might decide to work towards a \textit{minimum viable product} despite the challenges and work on \textit{recruiting} in the meanwhile.

\paragraph*{Technological tools} play a major role in facilitating team functionality. Communication channels such as Slack and WhatsApp and commit access to define team membership in one sense. If a team member is part of all shared online systems that enable them to contribute, that essentially signifies team membership. Furthermore, they store and communicate all group artifacts, and in some cases act semi-independently after programming and exert influence over the process. For example, software artifact production is facilitated through GitHub, which is a platform for storing and managing software source code. The platform facilitates source code peer review and can also perform automated quality control operations after each new commit. In one of the projects, the \textit{core group} has decided on automatic style guide checking through linter software and that each new commit has to be peer reviewed in a code review process in GitHub. Technically anyone can contribute new source code, even a complete newcomer. However, the contribution has to 1) match existing group goals, 2) pass automatic evaluation against pre-existing unit tests and the automated style guide, and 3) pass a peer review by more experienced contributors.
\begin{displayquote}
    ``At the moment, I would say it [our process] is leaning heavily towards being more technically evaluated. Every bit of code we write is reviewed. It's reviewed manually by another member of the team before it can get into the mainline development branch. It's also run-- We also have an automated test that run against those that check the code quality and just whether it works at a basic level.'' ---Interviewee D6
\end{displayquote}

By programming decisions and quality control checks into automated tools, the team codifies earlier decisions and \textit{invests power in them}. This is one of the reactions the team has taken to address the uncertain project stability and variable levels of commitment of each team member.

The approach to stakeholder involvement is a mix of open source approaches and a variant of agile. Like in many open source projects, all contributions and contributors are welcome. Unlike in large open source projects, the current group structure aims for closer, agile-style involvement and attempts to involve contributors in socially structured activities, such as the meet-ups, instead of just accepting software contributions. However, unlike in many agile processes such as SCRUM, there is no strict gate-keeping structure.
\begin{displayquote}
    ``I wouldn't call it sprints, I would say we work quite agile. I wouldn't call it sprints but we do kind of do like a week-by-week.'' ---Interviewee D5
\end{displayquote}

The group works with the intention of being part of a larger ecosystem of other, related groups, and sharing source code and adapting the shared solutions of other groups to local contexts. However, generalizing the solutions has proven to be challenging, considering the localized nature of problems they are addressing. They still participate in the network in the sense that they use open source -licensed software, openly publish their software artifacts, and publicize information about their solutions to other groups within the "Code for" network.

\subsection{Theoretical Coding: Abstracting the Development Process and Defining the Associations}

In this section, we present an abstracted form of the civic tech software development process, analyze how the unique operating environment and other characteristics affect it, and further refine relationships between the concepts and actors. 

\subsubsection{\textbf{Grassroots-driven Civic Technology Software Development Process}} \hspace*{\fill} \\
The civic technology group has a systematic development process with four major stages: Requirements and goals planning, architecture and data structure planning, iterative development that starts with prototyping, and evaluation. They are iterative and agile, with a possibility to return and adjust any stage after evaluation. We call it systematic and iterative because the group engaged in explicit meta-level process selection and development stage before engaging in actual software development. However, the process has some inflexibilities, because the group has less cohesion and face-to-face time. Due to those issues, the group relies more on the established processes and silos of responsibility.

\begin{figure*}
    \centering
    \includegraphics[width=\textwidth]{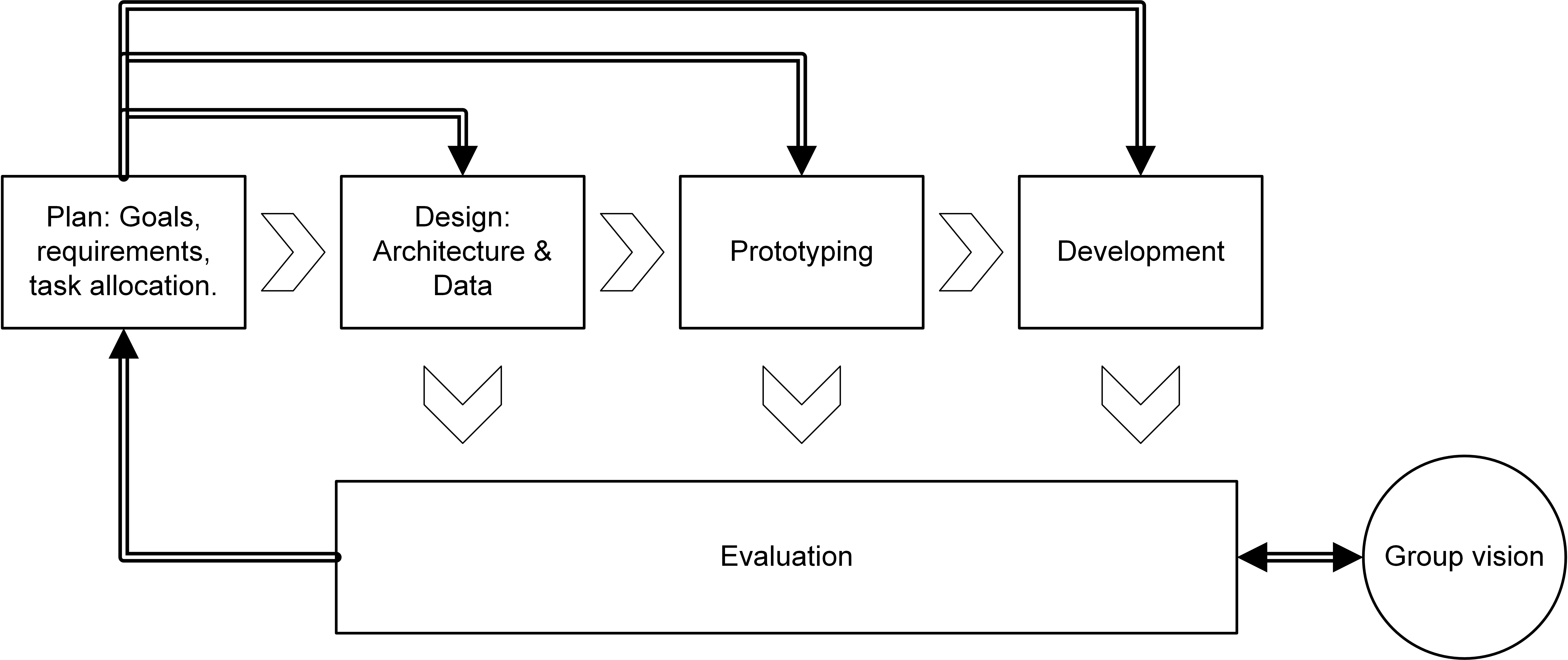}
    \caption{The iterative stages of the civic tech group's software development process}
    \label{fig:projectphases}
\end{figure*}

The development process and the flow between process stages are visualized in Fig.~\ref{fig:projectphases}. The large arrows denote transitions between project stages and the thin arrows denote flow of information between stages that do not have direct transitions. We detail each major stage in the group's development process further as follows.

\begin{itemize}
\item \textit{Goal and requirements planning, and task allocation.} The unique characteristics of this stage that there is no formal requirements engineering process. The planning starts as a brainstorming session that tries to fit the identified needs with participant motivations. The planning leads results in the \textit{group vision}, which is an informal agreement on what the project goal is and how it should be accomplished. From there each task is allocated a task lead, who takes responsibility for ensuring that each part of the project is finished. The tasks are distributed based on individual interests and competencies.
\item \textit{Architecture and data structure design.} The design process starts with architecture and data structure development. This is because the group aims for the parallel development of various software components that interact with standardized data structures and application programming interfaces. In order to enable parallel development, the group started both projects with architecture and data structure design.
\item \textit{Prototyping.} After planning and architecture design, the lead for each part designs a prototype that is evaluated against the group vision. The prototyping process and evaluation can also lead to new planning and task allocation if new needs or required features are identified. The use of prototypes is mainly internal and less used after the first design iterations.
\item \textit{Development.} The development process follows an iterative plan-consensus-action-evaluation loop that was described in Fig.~\ref{fig:designloop}. It should be noted that as many features as possible are developed in parallel, and evaluated individually. The development of quality control features such as unit tests and continuous integration pathways also proceeds simultaneously with development activities. Development is then continuously informed by automatic testing. The automatic testing feedback is personal, as opposed to group evaluation and peer review, which is performed together with the team.
\item \textit{Evaluation.} Evaluation is performed in two ways. The first one is the evaluation of the technical quality of new software contributions. It is carried out in GitHub and facilitated by GitHub's code review process, which allows other team members to evaluate summaries of automated testing results and review proposed changes to source code. This code review process can lead to acceptance or rejection of new contributions. The second type of evaluation is a shared, qualitative evaluation of overall software quality and functionality against the group vision. There are no formal quality requirement criteria, and the evaluation depends on the core group on agreeing whether the task outcomes meet the desired level of quality. This type of evaluation leads back to the re-negotiation of short term goals and new task allocations. Some evaluations also affect group vision.
\end{itemize}

The team members adopt different roles in the process, with the core group driving most of the processes forward and work being coordinated by a project manager if one has been selected. The mentor adopts a larger role in the planning and evaluation phases and acts as a kind of a sparring partner. The mentor never questions the goals themselves, allowing the team to be self-directed, but challenges the team in terms of rigor in development, realism in implementation goals, and evaluation methods.

More casual team members contribute to planning and development to the extent they can. However, more casual participants are invited to participate more, and one recurring meta-level discussion in the group is to how better give tasks to and benefit from the less committed participants.

\subsubsection{\textbf{Unique Characteristics Grassroots-driven Civic Technology Software Construction Process}} \hspace*{\fill} \\
Many of the group members have described methods they have adopted as agile. This is probably because agile is one of the most popular development methods in the local small to medium software companies, where most of the group's participating developers come from. However, there are also apparent differences to agile approaches. For example, before engaging in development activities, the team members to negotiate motivations, roles, and goals without much external input. Also, after basic structures and processes have been set up, there is an unwillingness to go back and change things, because that would involve a full re-negotiation many group structures.

In both projects the major effort was to create a \textit{minimum viable product} that would \textit{demonstrate group capabilities} in order to gain more \textit{credibility} in the eyes of \textit{external stakeholders}. For example, the mentor has mentioned that the current environment has an over-abundance of charitable and volunteering organizations. The drive to gain demonstrable outputs and more credibility was a driving meta-level goal. It could be said that the process to achieve a minimum viable product is an extended vetting process, with projects that are not able to reach this stage eventually starting to flounder and team members move to work on other projects or leave the group altogether. At the moment of writing, the Transparent Water project was closest to reaching the minimum viable product.

Two major characteristics of the process of an individual project, as presented in Fig.~\ref{fig:workloop}, are \textit{constant renegotiation of goals} and \textit{opportunistic evaluation}. Both \textit{monthly meetups} and \textit{weekly online standups} involve a constantly re-establishing consensus. For example, if an expert that was championing some particular technology leaves the group and later a person endorsing another technology joins the team, it is likely that the team will eventually shift to use the new one. This \textit{project drift} is related to \textit{opportunistic evaluation} and occasionally causes tension against the drive to achieve a \textit{minimum viable product}.

The major unique characteristics of the software construction process that are caused by volunteer-based workers, fluid group structure, drive to acquire more resources, opportunistic evaluation, and constant re-negotiation of goals are summarized below.

\begin{itemize}
    \item \textit{Internal stakeholders with development goals driven by team member motivations}. All stakeholders are considered internal to the project, in the sense that all concerned parties are invited to participate in the project. The current participants are mostly software engineers or from technical backgrounds, which causes the projects to be considered from a mostly technical background.
    \item \textit{Skeptical external stakeholders}. Grassroots-driven development does not have backing from large civic, academic or industrial organizations. Civic organizations can be unused or negative to civic participation, such as Irish Water in Dublin. One of the major challenges in new civic tech groups is to gain credibility and sell the idea of civic participation to civic organizations. For example, what is technically supposed to be open data, it is in practice closed from the stakeholders whom the civic authority deems "not worthy."
    \item \textit{Meritocratic involvement}. The more a participant does and the more they demonstrate competence, the more authority that participant gains. When a participant does more and contributes more, they acquire more responsibility and influence based on the respect and voluntary investment of authority from others. This is emphasized in requirements definition and evaluation. This leads to a \textit{emerging hierarchy}, who can steer the group through challenges. The more fluid the team becomes, the more a single person such as a project manager gains responsibility.
    \item \textit{Civic ghosting} and the challenge in harnessing casual participation. Some aspects of development are slowed by mismatches in communicated and intended participation. The unavoidable challenge caused by volunteering participants is that some participants simply disappear without any communication or \textit{ghost}.
    \item \textit{Opportunistic decision-making} and \textit{flexibility in evaluation} in the sense of skills, resources, and goals (``what we can do at the moment''). The team decides the next course of action partly based on previous goals, and partly based on reflection of team capabilities. Some essential tasks are delayed because they are seen to be unachievable for now. Furthermore, project stages stretch or shorten depending on how much time and resources the volunteer participants have currently.
    \item \textit{Uncertain project stage duration and project inertia}. Any project stage or iteration has no specific duration because the effort volunteers and other stakeholders can put in varies constantly. The project scope, duration, or goals are adjusted to reflect new resources and capabilities. Furthermore, the longer something has been a certain way, the more entrenched it becomes. This is can be a positive aspect, for example when the group has decided on well-functioning architecture and beneficial online tools. On the other hand, re-work and refactoring previous contributions are seen as less valuable, especially with the goal of achieving a minimum viable product. 
    \item \textit{Short-term perception}. The opportunistic decision-making, fluidity in team structure, and the drive to achieve a minimum viable product discourage long-term planning.
\end{itemize}

\section{Discussion}

The first question is whether a grassroots-driven civic tech group can set up a software engineering process. If we compare the discovered process against four fundamental characteristics of software engineering processes as defined by Sommerville \cite[p. 9]{sommerville_software_2011}: Software specification, development, validation, and evolution, we find out that the process has each of the characteristics. Even though the emphasis is on development and validation, both specification and evolution aspects are present.

If the group's process is a software engineering process, the question remains what kind of process it is, in relation to existing literature. Perng and Kitchin \cite{perng2018solutions} discussed Code for Ireland from a civic and cultural perspective and defined Code for Ireland as a civic hacking group, where "civic hacking binds together elements of civic innovation and computer hacking, with citizens quickly and collaboratively developing technological solutions" \cite[pp. 2]{perng2018solutions}. Based on our findings, we partly agree with this definition, though our discovery is that the group has evolved to a phase where they have systematized their process, and it has rigor beyond hacking.

Compared to existing software engineering processes defined in established literature \cite{sommerville_software_2011, dyba_empirical_2008}, the group's process is unique. It resembles agile methods most closely, conforming to the principles of the agile manifesto\footnote{http://agilemanifesto.org/}, with emphasis on individuals and interactions, working software, collaboration over contracts, and responding to change over following a plan. The significant difference is that the team is most commonly distributed, and has encountered difficulties often present in distributed development \cite{ramesh2006can}, such as communication challenges, lack of control, and lack of trust. We discuss the actions a distributed agile group can take to address these issues \cite{ramesh2006can} in the next subsection.

A taxonomy of agile principles by Conboy \cite{conboy2009agility} defines agility as creation of, proaction in advance of, reacting to, and learning from change. Agile methods should also contribute to perceived economy, quality, and simplicity. When using the taxonomy to evaluate the group's software engineering process in terms of agility, the group still has some issues in aspects of learning, creation, proactivity, and quality. Many of these challenges arise from the group's unique operating environment, but some could be addressed by improvements in the process. We discuss these in more detail in the next subsection.

When considering a wider context, the group had difficulties in external stakeholder engagement. Their issues are similar to ones discovered by Lee et al. \cite{lee2015open}. They discuss the civic app development efforts in the US, where programming contests and opening up of datasets were used to encourage the development of applications to address civic challenges. According to Lee et al., the first generation of efforts failed because of difficulties in value capture and lack of involvement within city departments. The motivation of citizens as participants was not fully understood and there was resistance from within the government to participate. The challenges Code for Ireland has faced are similar. While the participants in Code for Ireland are strongly motivated, the response to attempts of stakeholder engagement from civic authority stakeholders ranged from neutral to negative. In the second generation of civic app approaches \cite{lee2015open}, the engagement issues were addressed by legislation to force civic bodies to engage, providing support to bottom-up engaged communities, and bringing in entrepreneurs and venture capitalists to judging panels to evaluate most promising app ideas.

Additionally, when participants in Code for Ireland reflected on their processes, they identified issues with engaging newcomers and non-technically oriented participants. In these matters, lessons could be learned from the engagement aspects of academically-led co-creation and co-design approaches \cite{ferrario2014software, balestrini2017city, gooch2018amplifying}, innovation accelerators \cite{almirall2014open}, and open source software communities \cite{scacchi2006understanding}.

\subsection{Lessons Learned and Recommendations for Practise}

Perhaps the most significant challenges the group faced were due to the nature of the group and the operating environment. All participants were inhabitants of Dublin who wanted to effect change in their environment by the creation of digital services. This means that all development efforts were volunteered and any resources used were free or sponsored. The group was held back by a lack of engagement by the civic sector, which was a factor out of their control. However, some issues could be addressed by improvements in the development process and a more systematic approach to both citizen and civic stakeholder engagement. We list some of the main challenges and proposed remedies from literature as follows.

\begin{itemize}
\item \textbf{\textit{Civic ghosting and engaging newcomers}}. By using shared duties and methods such as pair programming, which has been found to increase agile learning in development processes \cite{conboy2009agility}, the group would address several issues: Harden any single feature under development against a participant disengaging from the group, increasing learning across silos, and helping new, inexperienced members to become fully contributing participants. Furthermore, having shared repositories, encouraging documentation and knowledge sharing, and concentrating on well-understood functionality can make achieving development goals easier \cite{ramesh2006can}.

\item \textbf{\textit{Transitioning from planning to development to release}}. Planning iterations \cite{ramesh2006can} and explicit retrospectives \cite{conboy2009agility} can help development groups to shift gears when transitioning between development stages. While the lack of resources can make every moment feel important, it is important to take a moment to re-evaluate processes and goals. At the same time, goals should be clear and documented in shared knowledge repositories. Documentation is more important in distributed development groups than in other agile approaches \cite{ramesh2006can}.

\item \textbf{\textit{Group cohesion and trust}}.  Without an external motivator such as income, the group is dependent on internal motivation to work together. As some of the participants were new to volunteering-based work, organizing and harnessing volunteer work was a challenge. In distributed teams, Ramesh et al. \cite{ramesh2006can} recommend building trust and community by distributed partners, sponsor visits, and building a cohesive team culture. Furthermore, clear and constant communication by easy to access methods, social media presence, or even a regular physical meeting location will promote team communication and identity.

\item \textbf{\textit{External mentoring and support}}. Paradoxically, a grassroots-driven approach does not ensure a wider citizen engagement. The group faced issues that have solutions proposed by researchers, especially in community engagement \cite{gooch2018amplifying} and hand-off \cite{balestrini2017city}. In this, the group could benefit from mentoring and lessons learned in the civic technology and HCI research community.

\item \textbf{\textit{Towards a Civic Tech Toolkit}}. Although several frameworks from projects in partnerships with citizens \cite{ferrario2014software,balestrini2017city,lee2015open} have outlined potential life-cycles to enhance social technology development. Their application seems to be currently limited to academic contexts. The creation of a civic tech toolbox similar to the current toolkits for orchestrating citizen sensing\footnote{Making Sense Toolkit: \url{http://making-sense.eu/publication_categories/toolkit}} or hackathons\footnote{The Hackathon Toolbox: \url{https://www.thecodeship.com/general/hackathon-toolbox-essential-tools-practices}} could help to solve this issue by allowing civic groups to discover, use, re-use and share content and knowledge relevant for their causes. 
\end{itemize}

While the group cannot address the lack of engagement from civic organizations, Lee et al. \cite{lee2015open} have listed actions civic organizations have taken that allowed grassroots civic technology groups to thrive. They included (1) legislation to force civic bodies to publish data in a timely manner, (2) publishing problem statements by cities to direct developer attention, (3) stronger management and direct coordination by city administrations, and (4) common app and crowd-sourced data repositories, along with bottom-up engaged communities.

\section{Conclusions}

In this paper, we investigated and described the software engineering process of a grassroots-driven civic technology group, and presented how the process resembles and diverges from agile approaches. Our preliminary findings show that such groups are capable of setting up systematic software engineering processes that address software specification, development, validation, and evolution. However, while they are able to deliver software according to self-specified quality standards, the group has challenges in requirements specification, stakeholder engagement, and reorienting from development to product delivery. The challenges they face are due to the civic environment, nature of participants, and issues in distributed development processes.

Grassroots-driven civic technology groups can deliver software, but our findings show that even if they have software engineering professionals as volunteers, they could benefit from improvements to their process. For example, they could utilize an existing community engagement process \cite{balestrini2017city,ferrario2014software,gooch2018amplifying}, and address issues in distributed development \cite{ramesh2006can} and agility \cite{conboy2009agility}. In this, the group could benefit from being mentored by a more experienced civic technology organization, even though that can increase the risk of the group becoming dependent on the mentor. However, many of the challenges the group faced were because of civic organizations and lack of support. If these were addressed from the civic side, such as in the US \cite{lee2015open}, it would allow an increased number of grassroots-driven civic technology groups to thrive.

We extended the current body of knowledge by presenting a case study of how a grassroots-driven civic tech software process emerges in the absence of a major academic, civic or industrial stakeholder. The way grassroots-driven civic technology group evolves is unique from existing top-down, orchestrated frameworks and will help both understand and better involve civic volunteer stakeholders in future efforts. As an additional contribution, this study demonstrates the suitability of interpretive case studies to inductively generate knowledge surrounding software creation processes.

We acknowledge a potential limitation for this study in the use of only a single case in the study. While generalized outcomes from one case study are achievable \cite{Urquhart2010Putting,yin2009case,eisenhardt1989building}, the debate is still ongoing. However, some case study methodologists argue that multiple case study is a more viable option to provide stronger assertions \cite{stake2013multiple}. Our future research will involve more cases. As a result, we expect further aspects and stronger assertions to emerge when conducting analysis across cases.

\section*{Acknowledgements}
The work of the first author was supported by the Ulla Tuominen Foundation. This work was supported, in part, by Science Foundation Ireland grant 13/RC/2094. We thank Lero, the Irish Software Research Centre, for their support.

\FloatBarrier

\bibliographystyle{IEEEtranN}
\bibliography{refs/knutas-zotero-export-1,refs/references-knutas-2,refs/palacin_refs,refs/giovannirefs,refs/vikitable}

\end{document}